\documentclass[%
 aip,
 amsmath,amssymb,
 reprint,%
]{revtex4-1}

\usepackage{graphicx}
\usepackage{dcolumn}
\usepackage{bm}

\usepackage[utf8]{inputenc}
\usepackage[T1]{fontenc}
\usepackage{mathptmx}

\begin{document}

\preprint{AIP/123-QED}

\title{Dynamic tuning of near-field thermal rectification using reconfigurable and phase-transition metamaterials}

\author{Fangqi Chen}
\affiliation{Department of Mechanical and Industrial Engineering, Northeastern University, Boston, MA, 02115, USA}
\author{Xiaojie Liu}
\affiliation{Department of Mechanical and Industrial Engineering, Northeastern University, Boston, MA, 02115, USA}
\author{Yanpei Tian}
\affiliation{Department of Mechanical and Industrial Engineering, Northeastern University, Boston, MA, 02115, USA}
\author{Yi Zheng}

 \email{y.zheng@northeastern.edu.}
 
\affiliation{Department of Mechanical and Industrial Engineering, Northeastern University, Boston, MA, 02115, USA}
\affiliation{Department of Electrical and Computer Engineering, Northeastern University, Boston, MA, 02115, USA}

%
%
\date{\today}

\begin{abstract}
Taking advantage of phase-transition and reconfigurable metamaterials, dynamic control of nanoscale thermal modulation can be achieved through the near-field radiative thermal rectification devices. 
This work proposes an active-tuning near-field thermal rectifier using reconfigurable phase-transition metamaterials.
The rectifier has two terminals separated by vacuum, working under a controllable operational temperature around the critical temperature of the phase-transition material VO$_2$. 
One of the terminals is a stretchable structure made of PDMS thin film and grating consisting of various types of phase-transition material. 
The effects of various inclusion forms and all the related geometric parameters are well analyzed.
The controllable nanoscale thermal modulation can be achieved and the ultrahigh rectification ratios of 23.7 and 19.8, the highest values ever predicted, can be obtained for two deformation scenarios, respectively. 
It will shed light on the dynamic tuning of small-scale thermal transport and light manipulation.
\end{abstract}

\maketitle

Active control of heat transfer is an important and challenging topic in thermal engineering and it attracts much attention recently. The thermal rectifiers based on conduction mechanism have been presented in several papers. \cite{kobayashi2009oxide,takeuchi2012improvement} However, because of the presence of Kapitza resistances and the speed of phonons, the performance of these types of thermal diodes has been restricted. \cite{ben2015contactless,ben2014near} Alternatively, thermal rectification based on radiative heat transfer is proposed. Some works have been conducted in both far-field \cite{ito2014experimental,van2012emissivity,ben2013phase,nefzaoui2014radiative,kasali2020optimization,ghanekar2017high,audhkhasi2019design} and near-field limits \cite{yang2013radiation,ito2017dynamic,otey2010thermal,basu2011near,van2011fast,van2011phonon,van2012tuning,huang2013thermal,yang2015vacuum,wang2013thermal,biehs2011modulation,ghanekar2018strain,xu2018surface,ghanekar2016high,didari2018near,shen2018high} experimentally \cite{ito2017dynamic,ito2014experimental} and theoretically.\cite{huang2013thermal,yang2015vacuum,audhkhasi2019design} Extensive studies have realized thermal rectification by utilizing polar materials\cite{otey2010thermal,basu2011near,wang2013thermal,didari2018near} and phase-transition materials.\cite{van2012tuning,ben2013phase,huang2013thermal,yang2013radiation,ito2014experimental} 

In the near-field regime, the radiative heat transfer exceeds the blackbody limit by several orders of magnitude because of the tunneling of evanescent waves and coupling of surface phonon or plasmon polaritons, thus outstanding thermal rectification can be realized. Thermal diode is one of the thermal rectifiers that modulate thermal transport: the heat flux in one direction is much larger than the one in the opposite direction depending on the temperature gradient. To evaluate the performance of a thermal diode, the rectification ratio $R$ is defined as $R=(Q_F-Q_R)/Q_R$, where $Q_F$ and $Q_R$ refer to the forward and reverse heat fluxes, respectively.
Ito \emph{et al.} experimentally investigate the far-field radiative thermal rectifier utilizing the phase-transition material VO$_2$ and tune its operating temperature by doping tungsten.\cite{ito2014experimental} Yang \emph{et al.} show numerical results of thermal rectification based on VO$_2$ and discuss the influence of film thickness.\cite{yang2013radiation,yang2015vacuum} Huang \emph{et al.} combine two phase-change materials, VO$_2$ and LCSMO, to improve the thermal rectification ratio to a higher value of 7.7 according to the definition in this letter.\cite{huang2013thermal} Besides the plate-plate structure, some microstructures and nanostructures, like surface grating, are also introduced to further enhance the rectifier performance.\cite{audhkhasi2019design,didari2018near,shen2018high,jia2018far}
\begin{figure}[b]
\includegraphics[width=0.42\textwidth]{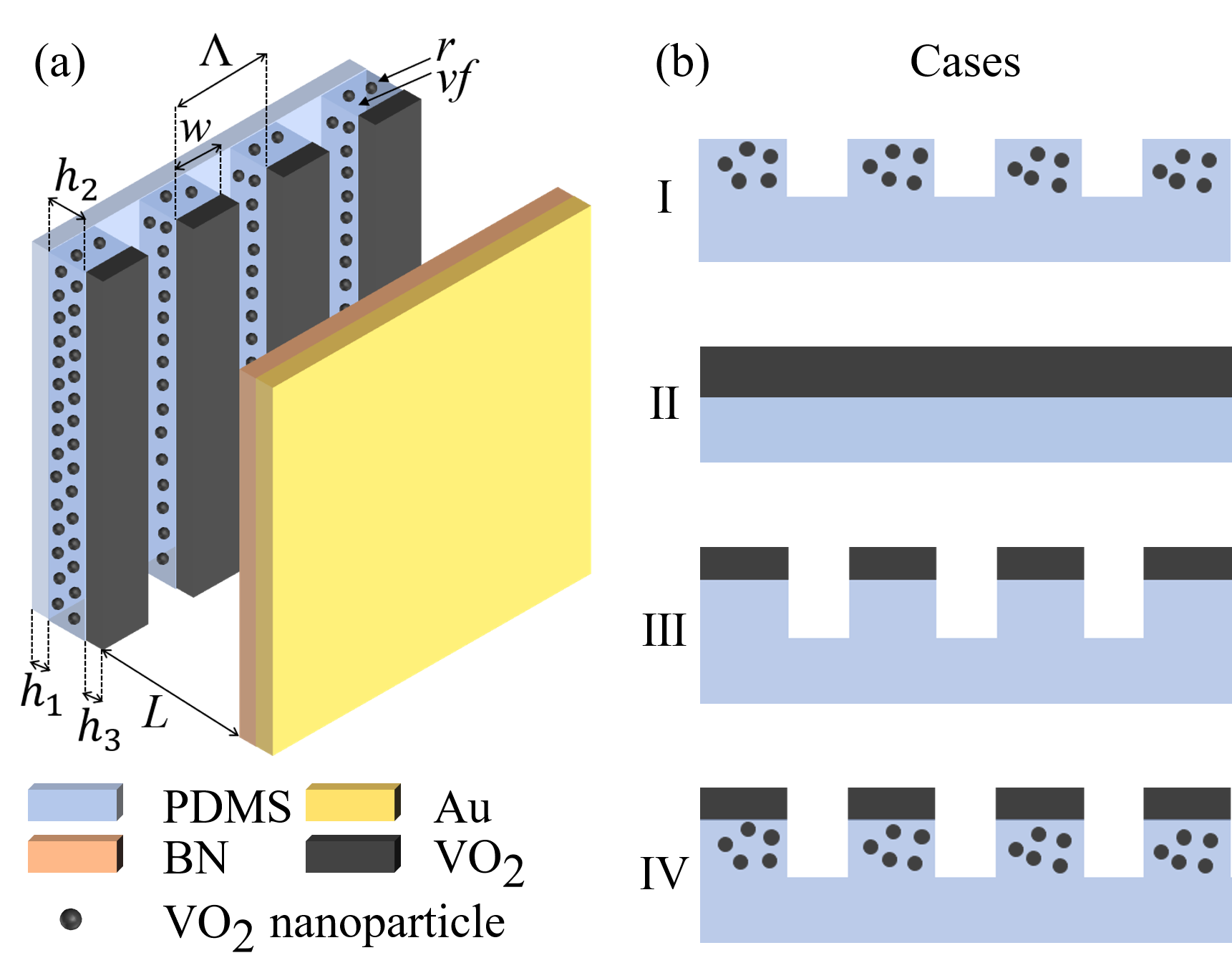}
\caption{\label{fig:fig1} (a) Schematic of a reconfigurable near-field thermal diode using phase-transition metamaterials. (b) Four structural cases for the active terminal. PDMS film thickness $h_1=60$ nm for all four cases. (\uppercase\expandafter{\romannumeral1}) PDMS grating with thickness $h_2=100$ nm, period $\Lambda= 50$ nm, filling ratio $\phi=0.2$, particle radius $r=20$ nm, and volume fraction $f=0.3$. (\uppercase\expandafter{\romannumeral2}) VO$_2$ film with thickness $h_3=100$ nm. (\uppercase\expandafter{\romannumeral3}) $h_2=100$ nm, $h_3=260$ nm, $\Lambda$ and $\phi$ are the same as case \uppercase\expandafter{\romannumeral1}. (\uppercase\expandafter{\romannumeral4}) $h_2$ and $h_3$ are the same as case \uppercase\expandafter{\romannumeral3}, $r$, $f$, $\Lambda$ and $\phi$ are the same as case \uppercase\expandafter{\romannumeral1}.}
\end{figure}

The phase-transition materials modulate the heat transfer due to their temperature-dependent thermal and optical properties, which renders a phase-transition based thermal device must operate around its critical temperature of phase change, restricting its range of applications. Besides the temperature dependence, the modulation of thermal transport can also be achieved by changing the configuration of a thermal device when subjected to mechanical strain. Biehs \emph{et al.} present a theoretical study that the twisting angle between two grating structures can modulate the net heat flux up to 90\% at room temperature.\cite{biehs2011modulation} Ghanekar \emph{et al.} propose a near-field thermal modulator exhibiting sensitivity to mechanical strain. \cite{ghanekar2018strain} Liu \emph{et al.} demonstrate a non-contact thermal modulator based on the mechanical rotation and a modulation contrast greater than 5 can be achieved.\cite{liu2017pattern} Though the studies on thermal rectifier driven by mechanical force are not much, reconfigurable metamaterials that can dynamically manipulate electromagnetic properties have aroused a lot of attention recently,\cite{lee2012reversibly,pryce2010highly,zheludev2016reconfigurable} paving the way for a deeper study on nanoscale radiative thermal rectification.

Here, we combine the comprehensive effects of phase transition and reconfigurable structures and propose a near-field stretchable radiative thermal diode based on the phase-transition material VO$_2$ and the soft host material PDMS with a tunable rectification. A comprehensive design is shown in Fig. \ref{fig:fig1}(a) and it has two terminals separated by a distance in nanoscale, which is less than the thermal wavelength of interest. The passive terminal is composed of 1 $\mu$m BN layer on top of 1 $\mu$m gold layer. The other terminal is referred to as the active terminal. The VO$_2$ film is deposited on top of a stretchable PDMS layer. In addition, VO$_2$ nanoparticles will be doped inside the PDMS grating structure. In Fig. \ref{fig:fig1}(b), we show four structural cases of interest for the active terminal while the passive terminal remains the same: (\uppercase\expandafter{\romannumeral1}) PDMS grating doped with VO$_2$ nanoparticles without VO$_2$ film deposited on top, (\uppercase\expandafter{\romannumeral2}) a thin VO$_2$ film, (\uppercase\expandafter{\romannumeral3}) VO$_2$ film deposited on top of the PDMS grating, (\uppercase\expandafter{\romannumeral4}) VO$_2$ film deposited on top of the PDMS grating, which is doped with VO$_2$ nanoparticles. The substrate for all the four cases is a 60 nm PDMS layer. For all the concepts, the temperature of the active terminal is set as $T_1 = 341$ K + $\Delta T$. The passive terminal has its temperature $T_2 = 341$ K $-$ $\Delta T$. Here, 341 K is the critical temperature of VO$_2$. When $T_1>T_2$ (referred to as forward bias), VO$_2$ is in metallic phase; when $T_1<T_2$ (reverse bias), VO$_2$ is in insulator phase with its optical axis normal to the surface. \cite{yang2013radiation}

\begin{figure}

\includegraphics[width=0.45\textwidth]{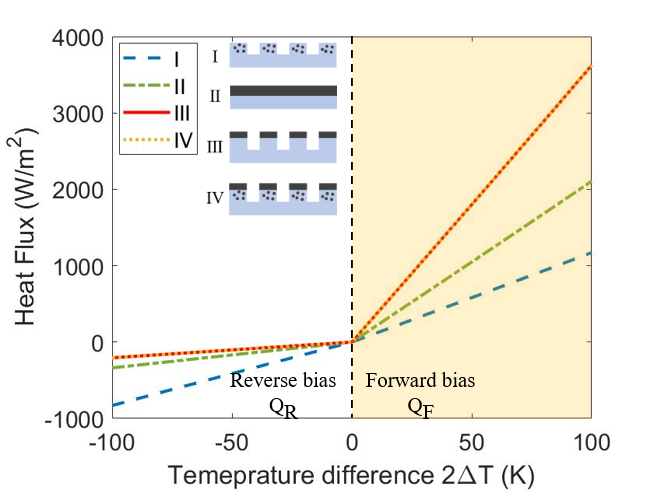}
\caption{\label{fig:fig2} Forward and reverse radiative heat fluxes ($Q_F$ and $Q_R$) versus the temperature difference between active and passive terminals at a 100 nm separation for four different cases.}
\end{figure}

The expressions for the radiative heat fluxes across the near-field thermal diode are obtained through the dyadic Green's function formalism.\cite{narayanaswamy2014green} The heat flux between planer objects can be calculated by
\begin{equation}
Q=\int_{0}^{\infty} \frac{d \omega}{2 \pi}\left[\Theta\left(\omega, T_{1}\right)-\Theta\left(\omega, T_{2}\right)\right] \int_{0}^{\infty} \frac{k_{\parallel} d k_{\parallel}}{2 \pi} Z \left(\omega, k_{\parallel}\right),
\end{equation}
where $T_1$ and $T_2$ are the temperatures of two objects, respectively. $\Theta(\omega, T)=(\hbar \omega / 2) \operatorname{coth}\left(\hbar \omega / 2 k_{B} T\right)$ is the energy of the harmonic oscillator. $\int_{0}^{\infty} \frac{k_{\parallel} d k_{\parallel}}{2 \pi} Z\left(\omega, k_{\parallel}\right)$ is known as the spectral transmissivity in radiative transfer between media 1 and 2 with gap $L$,
where $k_{\parallel}$ is the parallel component of wavevector and $Z\left(\omega, k_{\parallel}\right)$ is known as the energy transmission coefficient.
For our proposed 1-D grating structure of PDMS and VO$_2$, the second order approximation of the effective medium theory is used to obtain the effective dielectric properties.\cite{ghanekar2018strain} 

Effective medium approximation is only valid as the grating period $\Lambda$ is much less than the wavelength $\lambda$. Besides, the prerequisite for using the approximation in near-field radiative heat transfer is that the gap $L$ between the two objects is larger than the grating period. In this letter, the grating period (50 nm) is much less than the thermal wavelength around 341 K ($\sim$ 8.5 $\mu$m) and is smaller than the gap (100 nm) as well. Therefore, all these criteria are satisfied. In order to calculate the effective dielectric function of a composite medium containing nanoparticles in a host material, the Clausius-Mossotti equation is applied.\cite{ghanekar2015role}
\begin{figure}[b]

\includegraphics[width=0.45\textwidth]{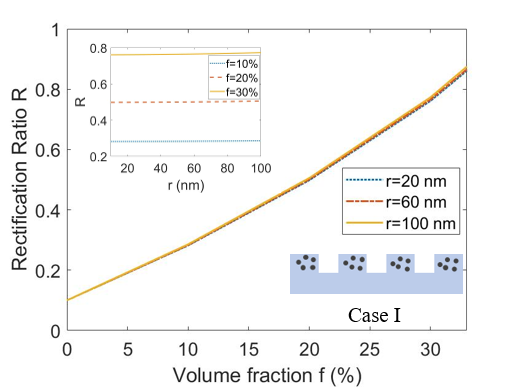}
\caption{\label{fig:fig3} The effects of volume fraction $f$ and radius $r$ (inset) of the VO$_2$ nanoparticle inclusions (case \uppercase\expandafter{\romannumeral1}) in a PDMS structure on thermal rectification ratio.}
\end{figure}

\begin{figure*}

\includegraphics[width=\textwidth]{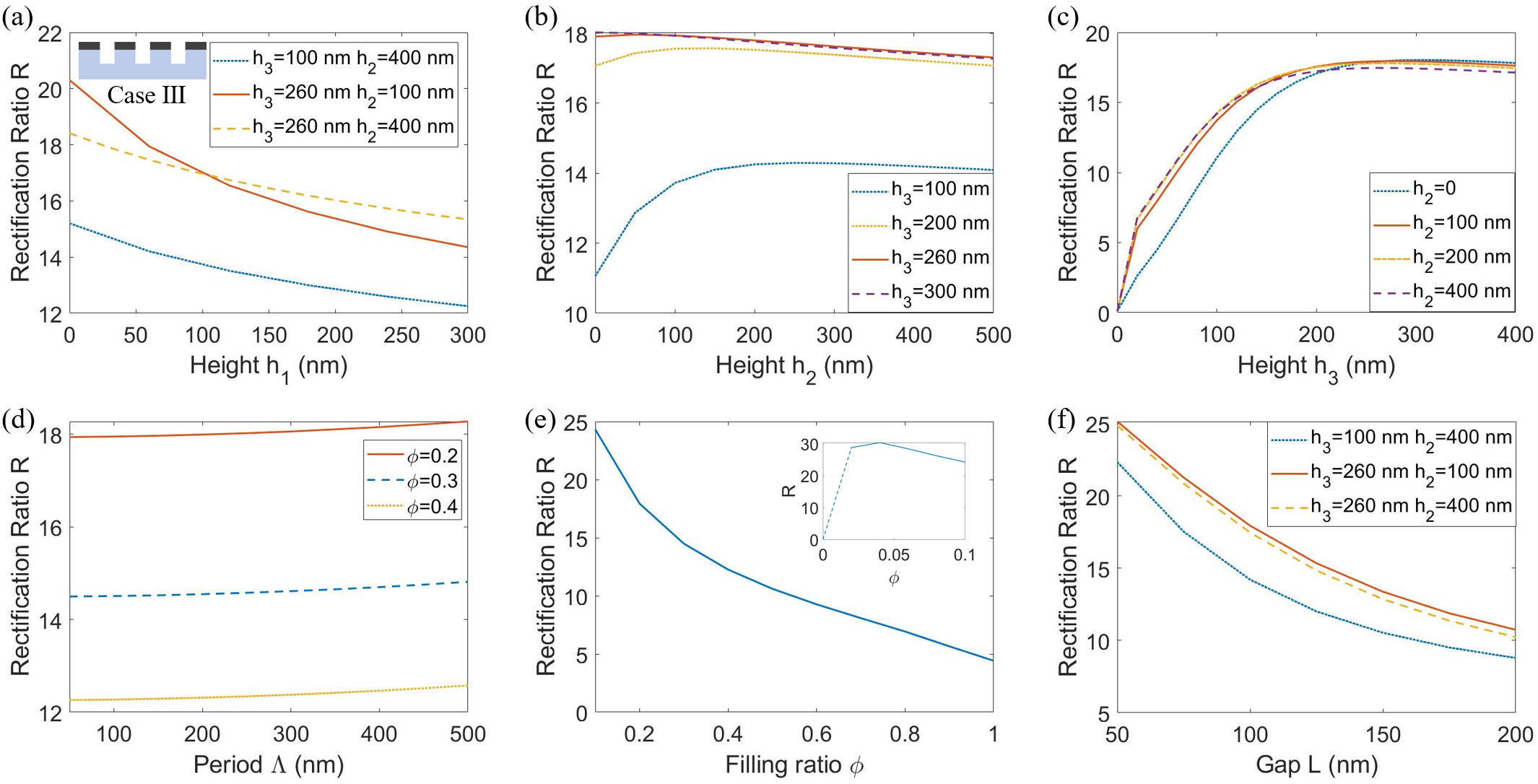}
\caption{\label{fig:fig4} The effects of geometric parameters on the rectification ratio $R$ for case \uppercase\expandafter{\romannumeral3}: (a) height $h_1$, (b) height $h_2$, (c) height $h_3$, (d) period $\Lambda$, (e) filling ratio $\phi$, and (f) gap $L$. The inset of figure (e) shows a trend of rectification values at small filling ratios.}
\end{figure*}

When the temperature of VO$_2$ is lower than its critical temperature (341 K), it is in the anisotropic insulator state. We can use the classical oscillator formula $\varepsilon(\omega)=\varepsilon_{\infty}+\sum_{i=1}^{N} \frac{S_{i} \omega_{i}^{2}}{\omega_{i}^{2}-j \gamma_{i} \omega-\omega^{2}}$ to determine  dielectric functions of the ordinary mode $\varepsilon_{O}$ and the extraordinary mode $\varepsilon_{E}$. The experimental values for calculation are given in the work of Barker \emph{et al.}.\cite{barker1966infrared}
As the temperature is above 341 K, VO$_2$ is in the isotropic metallic state. The Drude model is used to calculate the dielectric function $\varepsilon(\omega)=\frac{-\omega_{p}^{2} \varepsilon_{\infty}}{\omega^{2}-j \omega \Gamma}$. 
The dielectric function of gold can be found in the study by Johnson and Christy.\cite{johnson1972optical}  Querry \emph{et al.} provide the refractive indices of PDMS \cite{querry1987optical} and that of BN is given by Palik.\cite{palik1998handbook}

To give a brief evaluation of the thermal rectification performance of the four active terminal structures proposed in Fig. \ref{fig:fig1}(b), the heat flux dependence on the temperature difference is plotted in Fig. \ref{fig:fig2}. The right side of Fig. \ref{fig:fig2} (with yellow background) is referred to as the forward bias ($\Delta T>0$) with heat flux $Q_F$ and left side is referred to as the reverse bias ($\Delta T<0$) with heat flux $Q_R$. It can be seen that $Q_F$ is much larger than $Q_R$, showing an obvious thermal-diode feature, especially for the grating structures with VO$_2$ film deposited (case \uppercase\expandafter{\romannumeral3} and \uppercase\expandafter{\romannumeral4}). They have much better thermal rectification performances than the thin film structure (case \uppercase\expandafter{\romannumeral2}). Besides, the inclusion of VO$_2$ nanoparticles doesn't play an important role in heat modulation, which can be concluded from the following two aspects. First, case \uppercase\expandafter{\romannumeral1} exhibits the weakest thermal rectification among the four cases. Second, curves for case \uppercase\expandafter{\romannumeral3} and \uppercase\expandafter{\romannumeral4} almost overlap, showing little impact from the nanoparticle inclusion.

The effect of the nanoparticle inclusion is discussed in more details based on case \uppercase\expandafter{\romannumeral1}, as nanoparticle is the only inclusion form of the phase-transition material in this case. The rectification ratio is calculated at $T_1=331$ K for reverse bias and 351 K for forward bias, and $T_2=341$ K. The grating height of PDMS here is 400 nm. It can be observed from Fig. \ref{fig:fig3} that the rectification ratio increases monotonically with volume fraction, but it is quite limited. Due to the Maxwell-Garnett-Mie theory, the volume fraction threshold value is around 33\%, though higher volume fraction of nanoparticles is not feasible in real case since nanoparticles are easy to aggregate. If high volume fraction of VO$_2$ is needed, VO$_2$ can be involved as the host material rather than doped nanoparticles. Therefore, the doping process has little meaning. As we change the radius $r$, rectification ratio $R$ almost remains the same, so it can be concluded that radius is not a crucial factor in the thermal rectification. In one word, the nanoparticle is not a good inclusion form of phase-transition materials to realize a sharp thermal rectification, so in this letter case \uppercase\expandafter{\romannumeral3} is chosen as
an ideal structure and VO$_2$ is involved in the diode in the form of thin films.

The effects of all the related geometric parameters on the variation of rectification ratio are analyzed for case \uppercase\expandafter{\romannumeral3}. The parameters include the height of the PDMS substrate $h_1$, the height of PDMS grating $h_2$, the thickness of VO$_2$ film $h_3$, period $\Lambda$, filling ratio $\phi$ and separation distance $L$. Fig. \ref{fig:fig4}(a) shows a larger rectification ratio can be obtained at smaller $h_1$ in three given combinations of $h_2$ and $h_3$. Thinner PDMS substrate is preferred but a feasible value should be chosen for a practical device. As long as $h_3$ is thick enough ($>200$ nm), the rectification ratio remains quite stable above 17, while $h_2$ has little impact on the rectifier, which can be observed in both Fig. \ref{fig:fig4}(b) and (c). Here, $h_2$ is introduced in the diode mainly for an easy stretching rather than enhancing rectification. There is an optimal value for $h_3$, beyond which $R$ will decrease. The period is not a crucial parameter compared with filling ratio since $\Lambda$ is much smaller than the dominant thermal wavelength. When the filling ratio decreases, it optimizes the rectification ratio significantly. It is found in the inset of Fig. \ref{fig:fig4}(e) that $R$ rises up to 30 when the filling ratio $\phi$ is 0.04, impractically small, and goes down to zero when $\phi=0$, though such small filling ratio cannot be realized actually. Fig. \ref{fig:fig4}(f) displays the dependence of rectification ratio on the separation distance. It is easy to understand that $R$ increases at a smaller gap because of the strong variation of the intensity of evanescent waves, which is the fundamental mechanism of an ultrahigh near-field thermal rectification.\cite{huang2013thermal} The effect of evanescent waves become less dominant when gap $L$ increases, and it is negligible in the far-field thermal radiation. Based on the 
complete analysis above and considering the condition of the effective medium approximation ($L > \Lambda$), a set of optimal parameters is determined: $h_1=60$ nm, $h_2=100$ nm, $h_3=260$ nm, $\Lambda=50$ nm, $\phi=0.2$, $L=100$ nm, and an ultrahigh rectification ratio of 18 can be obtained.

More significantly in this letter, nanoscale thermal rectification is demonstrated as the active terminal is subjected to mechanical strain. Since VO$_2$ has a much larger Young's modulus than PDMS, it's assumed that VO$_2$ doesn't undergo any deformation. As in the original state, the active terminal has period $\Lambda$ and grating width $w$. Here, two scenarios for deformation are considered. The first one is an ideal scenario. The PDMS grating width $w$ remains unchanged as the period elongates from $\Lambda$ to $\Lambda + \Delta x$. However, from a more practical perspective, the PDMS grating must undergo deformation to some extent. It is assumed that the top width of the PDMS grating ($w_t$) remains unchanged and the bottom width ($w_b$) elongates in the same proportion to the substrate, which means the filling ratio is the same before and after deformation $(w/\Lambda=w_b/(\Lambda+\Delta x))$. The shape of the PDMS grating strip can be viewed as an isosceles trapezoid after deformation, which represents the real situation to some extent, though not exactly accurate, so it is known as the ideal actual scenario in this work. The change of $h_1$ and $h_2$ upon deformation is considered due to the fact that the Poisson's ratio for PDMS is 0.5, meaning the volume of PDMS remains constant during stretching or compression. 

\begin{figure}

\includegraphics[width=0.32\textwidth]{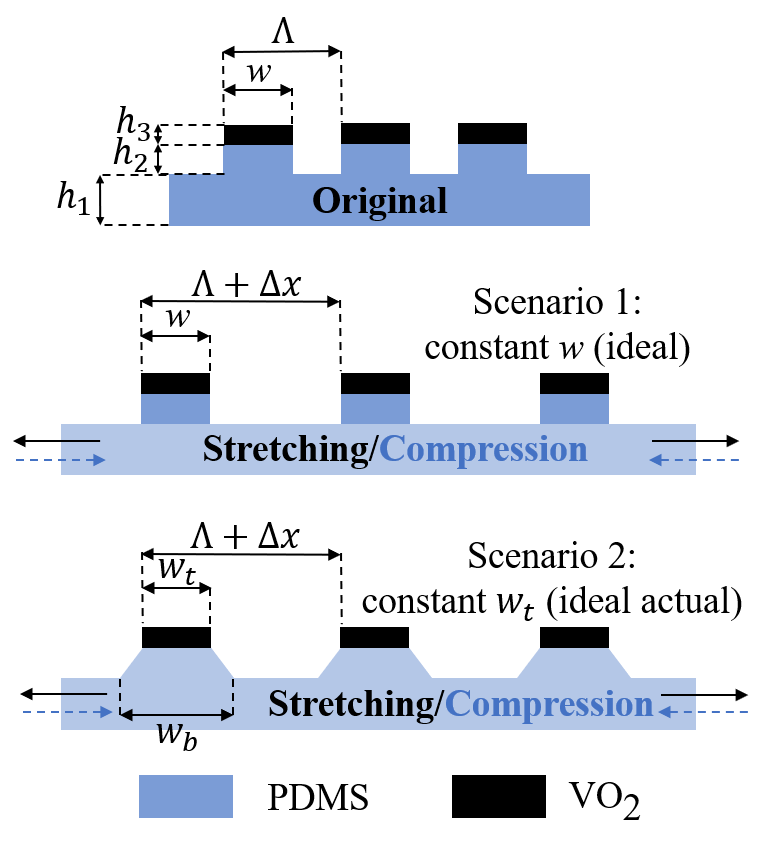}
\caption{\label{fig:fig5} Three states of the active terminal of  a reconfigurable near-field thermal diode. Top: the active terminal in its original state. Middle: scenario 1 (constant $w$) for an ideal stretching and compression deformation due to mechanical strain. Bottom: scenario 2 (constant $w_t$) for an ideal actual deformation due to mechanical strain.}
\end{figure}

\begin{figure}

\includegraphics[width=0.45\textwidth]{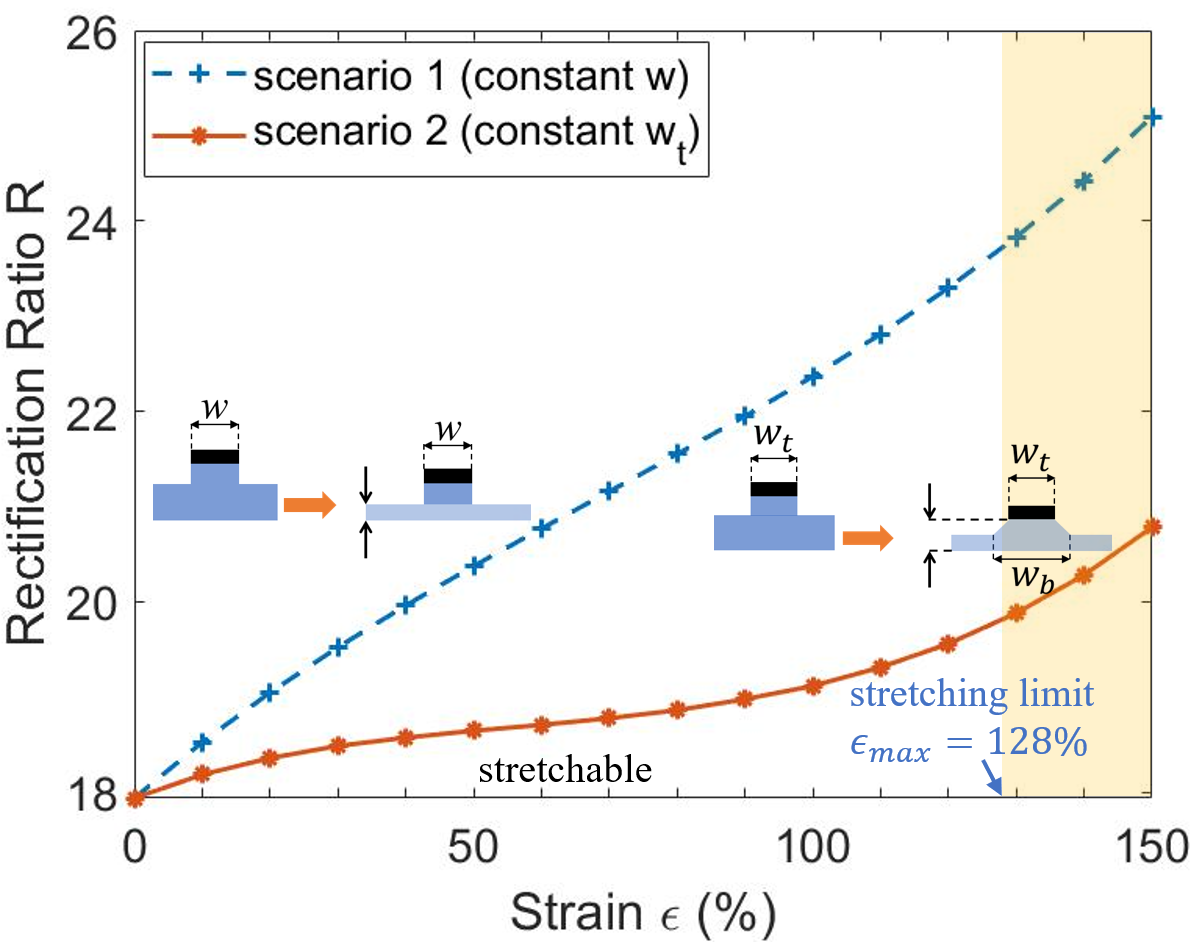}
\caption{\label{fig:fig6} Strain dependent thermal rectification  for two scenarios with the corresponding changes in heights ($\Delta h_1$ and $\Delta h_2$) while considering the Poisson's ratio of the soft host material PDMS in stretching.}
\end{figure}

To illustrate the stretching or compression process for the active terminal, the strain is defined as the change in period over the period of grating ($\Delta x/\Lambda$). For the deformation in scenario 2, the grating strip can be considered approximately as a composition of multiple layers of rectangular gratings, with increasing filling ratio from top to bottom. Here, 50 layers are used in the calculation and it is enough to get converging solutions. As seen in Fig. \ref{fig:fig6}, the rectification ratio increases along with strain in both scenarios. It increases much slower for scenario 2 compared with scenario 1, and the reason is that the grating structure in scenario 1 tends to cover less space on the substrate, \emph{i.e.}, smaller filling ratio, upon deformation. It is concluded that a smaller filling ratio contributes to a larger rectification ratio. Using the data from Wu \emph{et al.},\cite{wu2018stretchable} the PDMS film will break at the strain of 128\%, which is set to be the stretching limit in this work. The area beyond the limit with yellow background cannot be reached actually. At the stretching limit, the rectification ratio for scenario 1 and 2 are 23.7 and 19.8, respectively. Above all, an increasing trend of radiative thermal rectification effect is observed when a reconfigurable near-field thermal diode undergoes deformation.

In conclusion, many studies have been conducted on realizing thermal rectification in both far-field and near-field radiation by utilizing phase transition or varying configurations. In this letter, a dynamic tuning near-field thermal diode using reconfigurable and phase-transition metamaterials is explored. It is designed in a nano-grating structure and works around the critical temperature of phase-transition material VO$_2$. The best inclusion forms of VO$_2$ in thermal diodes are studied and it is found inclusion as thin films outperforms nanoparticles. The geometric parameters of the thermal diode may play an important role in enhancing rectification, like the thickness of VO$_2$ film and the filling ratio of the grating. With the determined optimal parameters, an ultrahigh rectification ratio of 18 in the original state is obtained. Besides, the effect of mechanical strain on the rectifier is well analyzed. Two scenarios are considered for the deformation process and a rising trend of rectification ratio is presented upon deformation as well as the highest rectification ratio equals to 23.7 to date. More study can be carried out on designing structures more sensitive to mechanical strain and temperature gradient. This work verifies the possibility of improving thermal rectification through reconfigurable nanostructures utilizing phase-transition metamaterials. It sheds light on the high-performance thermal diodes and motivates promising applications in dynamic control of nanoscale thermal transport. More future work will be addressed in thermal rectification by radiative thermal diode, like studying alternative phase-transition and host materials, designing nanostructures of active/passive terminals and exploring other heat modulation mechanisms.

~\\
\indent This project was supported by the National Science Foundation through Grant No. 1941743.

~\\
\indent The data that support the findings of this study are available from the corresponding author
upon reasonable request.
\nocite{*}

\providecommand{\noopsort}[1]{}\providecommand{\singleletter}[1]{#1}%

\end{document}